\documentclass[ reprint,
 amsmath,amssymb,
 aps,
prf,
onecolumn,
]{revtex4}
\usepackage{graphicx} 
\newcommand{\mbs}[1]{\boldsymbol{#1}}

\begin{document}
\centerline{\Huge }
 \vspace{2cm} 
\centerline{\LARGE Material optimization of flexible blades for wind turbines.} 
 \vspace{.5cm} 
 \centerline{V. Cognet$^1$, S. Courrech du Pont$^2$, B. Thiria$^1$.} 
 \vspace{1cm} 
\centerline{\footnotesize $^1$: Laboratoire de Physique et Mécanique des Milieux Hétérogènes, PMMH UMR 7636 CNRS —}
\centerline{\footnotesize ESPCI PSL Research University — Univ. Paris-Diderot — Sorbonne Université, Paris, France.}
\centerline{\footnotesize Barre Cassan, Bât A, 1er étage, Case 18, 7 quai saint Bernard, 75005 Paris.}
 \vspace{.5cm}
 \centerline{\footnotesize $^2$: Laboratoire Matière et Systèmes Complexes, UMR CNRS 7057, Université de Paris}
  \centerline{\footnotesize Bât. Condorcet, 10 rue Alice Domon et Léonie Duquet, 75013 Paris, France.}
 \vspace{2cm} 
\thispagestyle{empty}
\thispagestyle{empty}

\thispagestyle{empty}
\section*{Abstract}
Bioinspired flexible blades have been recently shown to significantly improve the versatility of horizontal-axis wind turbines, by widening their working range and increasing their efficiency. The aerodynamic and centrifugal forces bend the blade along its chord, varying the pitch angle by means of non-consuming mechanisms.
Here we introduce a general method based on a universal scaling, which finds the optimal soft materials for the blades to maximize the overall turbine efficiency or rotational power, for any required geometry of classical horizontal-axis turbines. The optimization problem, which depends on various parameters, such as the wind velocity, the rotation rate, the density, the rigidity and the geometry of the blade, is reduced to only two dimensionless parameters: the Cauchy number and the centrifugal number. The blade-element momentum theory is coupled to a torsion spring-based model for the blade deformation.
Taking into account realistic incoming wind-velocity distributions in the North Sea and a large wind-turbine geometry, we found a significant increase of the total harvested power, up to $+35\%$. In addition, the optimal soft material corresponding to the maximal efficiency over the entire working range for a given wind turbine geometry is, within the limits of small blade deformations, scale-independent. Thus experiments on small wind turbines are a possible way to determine the optimal soft materials for larger ones. These flexible blades are found to be between $5\%$ and $20\%$ lighter than the current rigid blades.
\section*{Keywords}
Wind energy, bio-inspiration, reconfiguration mechanism, optimization problem, energy saving, pitch control.
\clearpage
\section{Introduction}
In a recent paper \cite{cognet_bioinspired_2017}, we presented a new type of wind turbine with bioinspired flexible blades. These blades passively change their pitch angle with respect to the external wind conditions: the drag force and the lift force increase the pitch angle, while the centrifugal force decreases its absolute value. The optimal pitch angle distribution along the blade which maximizes the efficiency of the turbine varies with the working point (for example, with the incoming wind velocity or with the rotation rate). Changing the pitch angle by means of flexible blades is a way to follow the variation of this optimal pitch angle distribution and to increase tfhe versatility of the turbine. As a result, the efficiency was significantly improved (up to $+35\%$) over the entire working range of the turbine by tuning the density, the flexibility and the thickness of the flexible blade. Recently, these results and the order of magnitude increase in efficiency were confirmed on another small wind turbine with different experimental wind conditions \cite{macphee_performance__2018}.\\
Prior to this, a few studies had shown numerically \cite{krawczyk_fluid_2013, macphee_fluid-structure_2013} and experimentally \cite{macphee_experimental_2015} the interest of using soft homogeneous blades for a few specific sets of rigidities and wind velocities. All these studies were inspired by observations of various natural phenomena \cite{beyene_asfaw_constructal_2009}: reconfiguration of leaves in high winds decreases the drag force \cite{vogel_drag_1989, alben_drag_2002, gosselin_drag_2010}; flexible wings improve insect propulsion \cite{thiria_how_2010, ramananarivo_rather_2011}. Concerning propulsion, passive change of the twist of flexible blades is already used to improve the overall efficiency of multifunctional aircrafts (hover and forward flight) \cite{mohd-zawawi_study_2014, lv_performance_2015}.\\
Passive pitch control is a way to save weight and complexity in the rotor design \cite{ponta_adaptive-blade_2014}, compared to active pitch control methods \cite{dai_aerodynamic_2011, schubel_wind_2012, lachenal_review_2013} which are costly and energy consuming. As wind turbines are becoming increasingly larger, blades will likely be characterized by large displacements of the blade section \cite{ponta_effects_2016}. Alternative passive pitch control methods use anisotropic composite laminate to couple bending and torsion \cite{ponta_adaptive-blade_2014,karaolis_active_1988, lobitz_use_2001}.\\
This work introduces a general method based on a universal scaling aimed at finding the optimal soft materials for the blades maximizing either the efficiency or the rotational power of the turbine over its entire working range, for any required geometry of classical horizontal-axis turbines. Only two scale-independent parameters are necessary to find the optimal materials: the Cauchy number and the centrifugal number. The method is numerically applied to a commercial wind turbine of radius $R=8.5\ m$ described in \cite{burton_wind_2011}. First the model for the turbine with flexible blades is detailed. The Blade Element Momentum (BEM) theory is used to take into account aerodynamic profiles which strongly interact with the wind. This theory is coupled with a torsion spring based model for the blade deformation. The model can consider pre-twisted flexible blades and a radius-dependent chord. Secondly, results from this method are displayed and discussed for two different operating modes: constant wind velocity and constant rotation rate.\\
We show that for the two different operating modes, the overall efficiency and harvested power are increased, from a few to more than $100\%$. For wind velocity distributions corresponding to wind farms in the North Sea \cite{Coelingh_Analysis_1996}, the power gain goes up to $+35\%$. We believe that this method can be applied to any other operating modes. We find that the optimal soft material for a given turbine geometry is scale-independent, offering new perspective for wind tunnel experiments. This method was patented recently (publication number: WO2018/149895).

\section{Modeling}
\label{Modeling}

\begin{figure}[]
\centering
\includegraphics[height=7.8cm]{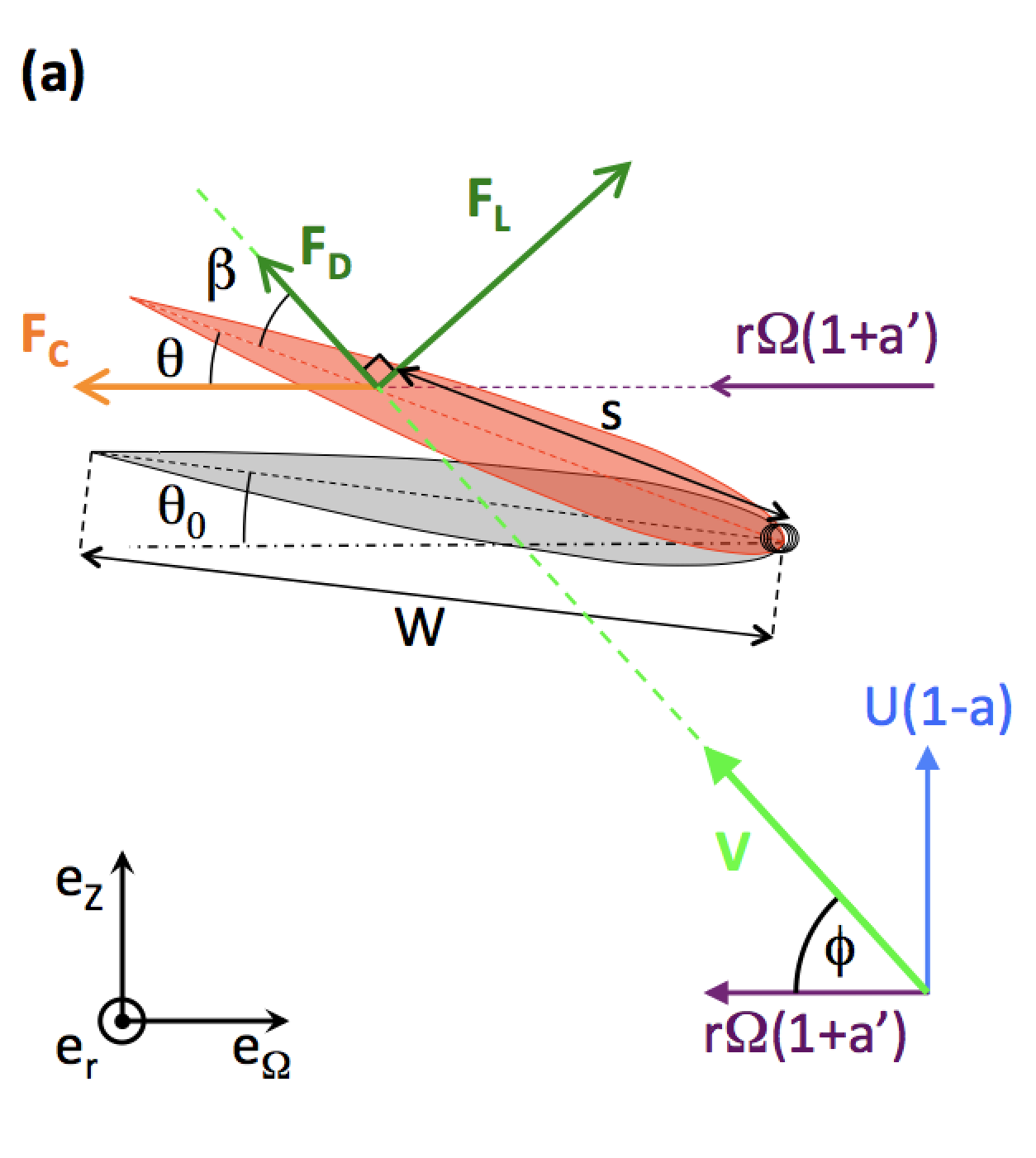} 
\includegraphics[height=7.8cm]{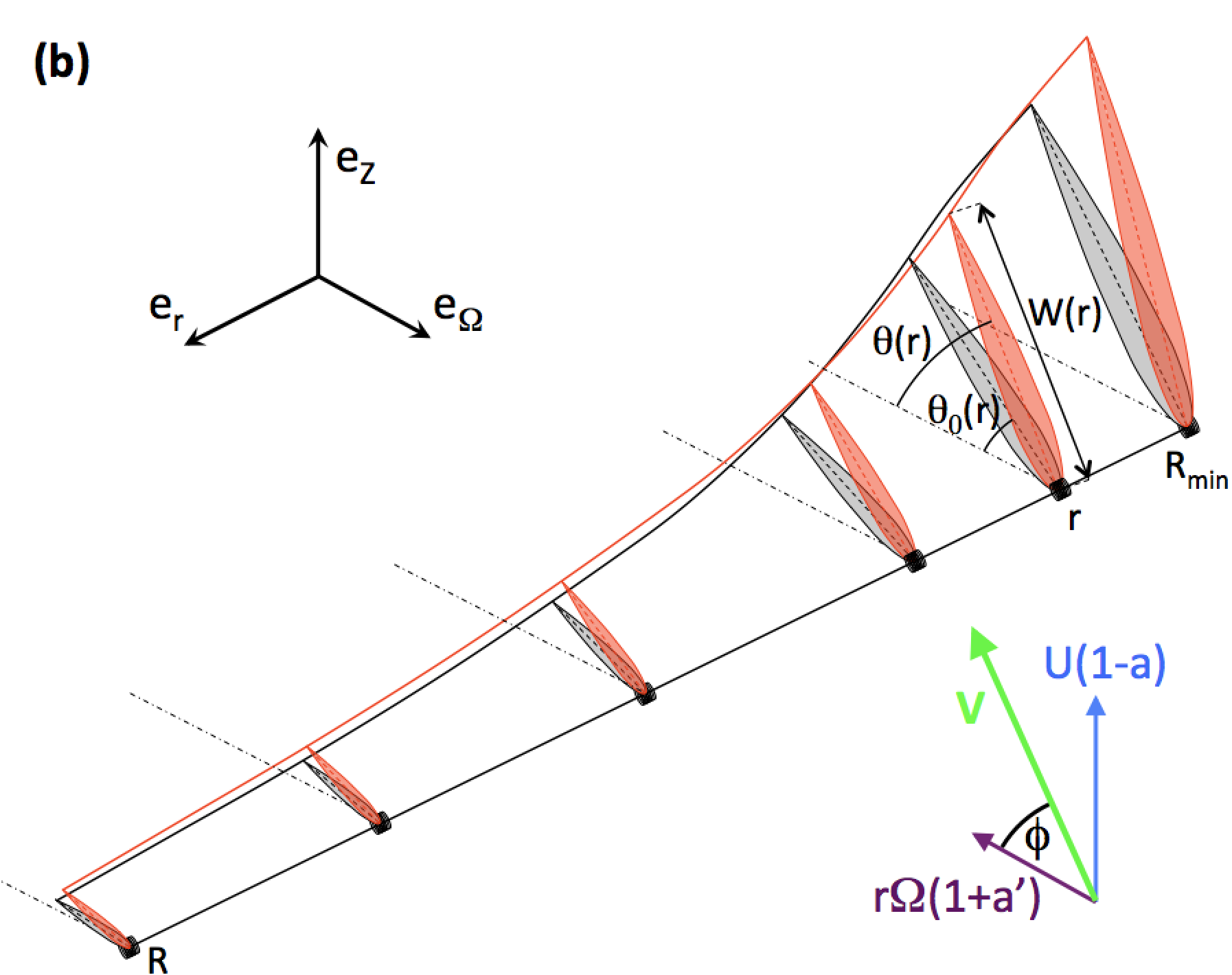} 
\caption{\footnotesize{(a) Sketch of the forces on the flexible blade section (in red) of chord $W$, located at a radius $r$ from the center. The lift force $\mbs{F_L}$ and the drag force $\mbs{F_D}$ tend to increase the pitch angle $\theta$ compared to its initial value $\theta_0$. The initial shape of the blade is in grey. The centrifugal force $\mbs{F_C}$ tends to reduce $| \theta |$. The deformation of the flexible blade section is modeled by a rotation around the torsion spring located at the leading edge. This rotation modifies the angle of attack, $\beta$, of the effective wind velocity $\mbs{V}$ on the blade section, which changes the efficiency $C_P$ of the wind turbine. (b) Sketch of different blade elements of the whole flexible blade (in red) and of its initial shape (in grey). The chord $W$ and the pitch angle $\theta$ vary with the radius $r$.}}
\label{Fig1}
\end{figure}
We present a model to evaluate the performance of an Horizontal-Axis Wind Turbine (HAWT) with rigid or flexible blades. We choose to use the BEM theory \cite{lanzafame_fluid_2007, Sanderse_Aerodynamics_2009}, a classical theoretical framework to study wind turbines aerodynamics. In this theory, the momentum variation of the air passing through the area swept by the blades is a function of the aerodynamic forces - the lift force $\mbs{F_L}$ and the drag force $\mbs{F_D}$ - applied on the blade sections (see \ref{AppendixBEM} for details). For each blade section located at a radius $r$ from the center of the rotational axis, the presence of the turbine modifies the local effective wind velocity $\mbs{V}(r)$, through the axial induction coefficient $a(r)$ and the tangential induction coefficient $a'(r)$, which varies between $0$ and $1$, (see figure \ref{Fig1}(a)). For a given blade section at radius $r$, the norm of the effective wind velocity is:
\begin{equation}
V = \sqrt{[U(1-a)]^2 + [r\Omega(1+a')]^2},
\label{Eq1}
\end{equation}
where $U$ is the incoming wind velocity, and $\Omega$ is the angular rotation rate. The efficiency $C_P$ of the turbine is defined as the ratio between the power harvested and the power of the wind going through the area swept by the blades:
\begin{align}
C_P=\frac{2C\Omega}{\rho U^3 \pi R^2}=\frac{2P}{\rho U^3 \pi R^2},
\label{Eq6}
\end{align}
where $C$ is the resistive torque applied by the generator, $\rho$ is the density of the air, and $P$ is the rotational power. In the following, we aim at maximizing the total efficiency of the wind turbine over its entire working range. The soft materials that attain this goal are referred to as optimal.\\
\noindent Commercial wind turbine blades are often twisted in order to widen the working range, especially to facilitate the starting stage. To take into account such geometries, the blade is modeled as a series of independent blade sections of chord $W(r)$. Every blade section located at a radius $r$ has its own initial pitch angle $\theta_0(r)$. In the case of a rigid blade, the effective wind velocity $\mbs{V}$ arrives on the blade section with an angle of attack $\beta(r)=\phi(r)-\theta_0(r)$, where $\phi(r)$ is the angle between $\mbs{V}$ and the rotation plane (see Fig. \ref{Fig1}).\\
\noindent During operation, the flexible blade section is subjected to different forces: on the one hand, the lift force $\mbs{F_L}$ and the drag force $\mbs{F_D}$ tend to increase the value of the pitch angle, while on the other hand, the centrifugal force $\mbs{F_C}$ tends to reduce the absolute value of the pitch angle (see Fig. \ref{Fig1}(a)). This passive variation of the pitch angle from its initial value $\theta_0$ to $\theta$ results in a change of $\beta$, and in a modification of the efficiency of the turbine \cite{cognet_bioinspired_2017}.\\
For the sake of simplicity, the deformation of the blade section is modeled by a rotation around the leading edge with a torsion spring, as shown on figure \ref{Fig1}(b). The blade element is assumed to be non-deformable, meaning that pure curvature effects are not studied. This assumption is acceptable only for small deformations. This point will be adressed in the discussion section.\\
On the blade section of width $dr$, aerodynamic and centrifugal forces respectively exert an aerodynamic torque $M_{aero}$ and a centrifugal torque $M_{cent}$ around the torsion spring with spring constant $K$. The difference between $\theta$ and $\theta_0$ is proportional to the total torque:
\begin{eqnarray}
   K(\theta-\theta_0) &=  & \int_0^W s \frac{\rho}{2} \left[ (U(1-a))^2 + (r\Omega (1+a'))^2 \right] \left(C_L(\beta) \cos(\beta) + C_D(\beta) \sin(\beta) \right)  \mathrm{dr}\ \mathrm{ds} \nonumber \\
&   & - \int_0^W s \rho_{blade} h \Omega^2 \sin(\theta) s \cos(\theta) \mathrm{dr}\ \mathrm{ds},
\label{Eq2}
\end{eqnarray}
\noindent where $s$ is the length between the torsion spring and the point considered on the blade (see Fig. \ref{Fig1}(a)); $C_L$ and $C_D$ are respectively the lift and drag coefficients, $\rho_{blade}$ is the density of the blade, and $h_0$ is the thickness of the blade contour. Since none of the parameters within the integrals vary with $s$, integration of equation \eqref{Eq2} leads to:
\begin{align}
\theta= \theta_0+ C_Y \left[ (1-a)^2 + (\lambda (1+a'))^2 \right] \left(C_L(\beta) \cos(\beta) + C_D(\beta) \sin(\beta) \right) - C_C \Lambda^2 \sin(\theta)  \cos(\theta),
\label{Eq3}
\end{align}
\noindent where $\beta=\phi-\theta$, with $\phi=\arctan{\left[(1-a)/ (\lambda(1+a')) \right]}$. $\lambda=r\Omega / U$ is the speed ratio at a radius $r$. In the case where $r=R$, the tip-speed ratio is referred to as $\Lambda$. $C_Y$ and $C_C$ are respectively referred to as the Cauchy number and the centrifugal number that represent the aerodynamical and the centrifugal intensities compared to the restoring force of the torsion spring, respectively.
\begin{align}
& C_Y =  \frac{\rho U^2 W^2 dr}{4K}, & C_C = \frac{\rho_{blade} U^2 h_0 W^3dr}{3R^2K}.
\label{Eq4}
\end{align}
The definitions of these dimensionless numbers have been given in the case where the flexible blade is modeled as an elastic beam \cite{cognet_bioinspired_2017}. The Cauchy number is also found in the studies of many reconfiguration mechanisms involving fluid-structure interaction \cite{gosselin_drag_2010, schouveiler_rolling_2006}. In equation \eqref{Eq3}, $a$, $a'$, $\lambda$ and $\beta$ depend on the radius $r$ and on the tip-speed ratio $\Lambda$. This means that the evolution of the pitch angle $\theta(r,\Lambda)$ is governed by only three parameters: the initial pitch angle distribution $\theta_0(r)$, the Cauchy number $C_Y$ and the centrifugal number $C_C$. The procedure will take advantage of this reduction of the number of parameters.\\ 
\noindent In the following, characteristic values of the wind turbine with flexible blades are plotted as a function of the input parameters $C_Y$ and $C_C$, ranging from 0 to around $0.025$. This range of values is especially small but corresponds, however, to significant pitch angle variations of up to $10$ degrees. This is because the effective aerodynamical and centrifugal intensities compared to the restoring force of the torsion spring are respectively $C_Y \left[ (1-a)^2 + (\lambda (1+a'))^2 \right]$ and $C_C \Lambda^2$, as shown in Eq. \eqref{Eq3}. These quantities are not constant along the working range, because they vary with $r$ and with $\Lambda$. We will see in the following that the values of these quantities can be larger than unity. The ratio between these two intensities predicts how the pitch angle evolves during operation. To evaluate the order of magnitude of the ratio, we set $a$ and $a'$ to zero:
\begin{align}
\frac{C_C \Lambda^2}{C_Y (1+\lambda^2)}=  \frac{4}{3} \frac{\rho_{blade}}{\rho} \frac{hW}{r^2} \frac{\lambda^2}{1+\lambda^2}.
\label{Eq5}
\end{align}
\noindent Equation \eqref{Eq5} shows that the ratio between the two intensities can be changed by carefully tuning the density and the geometry of the blade. It also points out that the centrifugal effect grows compared to the aerodynamic effect as the speed ratio $\lambda$ increases. Thus pitch angle decreases with $\lambda$.
\section{Method}
\label{Method.}
We present a method to determine the Young modulus $E$ and the density $\rho_{blade}$ of the optimal materials maximizing the total efficiency of the turbine. The method is applied to a commercial three-bladed wind turbine described in \cite{burton_wind_2011}. 
The chord varies linearly with the radius, from $W_{max}=1.085\ m$ at minimum radius $R_{min}=1.7\ m$, to $W_{min}=0.445\ m$ at maximum radius $R=8.5\ m$. The aerodynamic coefficients used are inspired from the classical NACA 0012 airfoil \cite{abbott_theory_1959}.\\
\noindent We first detail the three steps of the method in the case of a constant incoming wind velocity $U=6.9\ m.s^{-1}$. The value of $U$ is chosen such that the nominal point ($\Lambda^{opt}=7.8$) corresponds to a rotation rate $f=\frac{\Omega}{2\pi}$ equal to $1\ Hz$, in agreement with the description of the wind turbine given in \cite{burton_wind_2011}. This operating mode can be used in wind tunnel experiments: the angular rotation rate $\Omega$ decreases as the resistive torque $C$ increases. Equation \eqref{Eq3} shows that both aerodynamic forces and centrifugal force vary as $\Omega$ is changed, leading to an evolution of the balance between the two effects.\\
\begin{figure}[]
\centering
\includegraphics[width=1\linewidth]{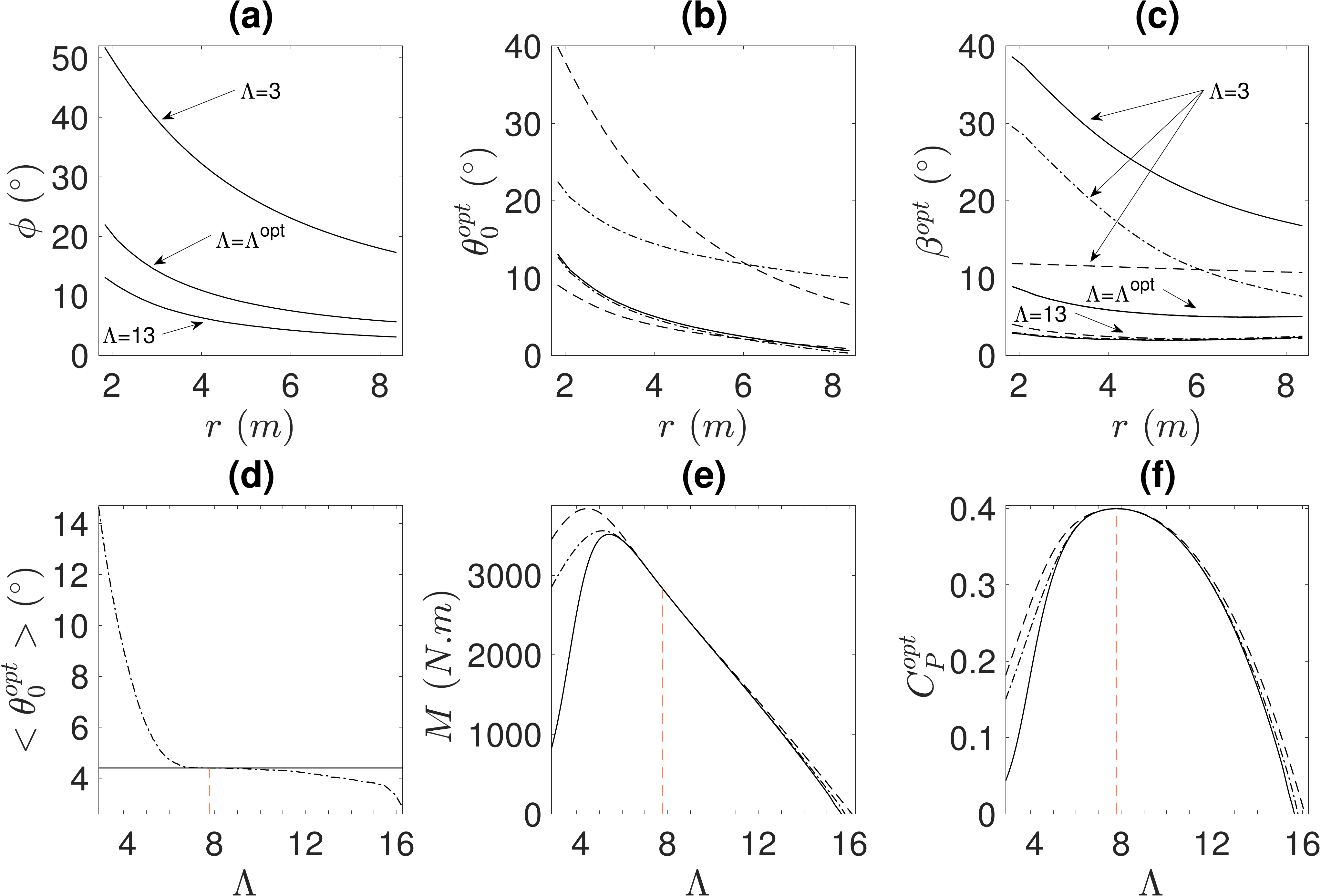} 
\caption{
\footnotesize{Evolution of the characteristic parameters of the wind turbine working at constant incoming wind velocity $U=6.9\ m.s^{-1}$, for three types of pitch angle variations: the reference case at optimal fixed pitch angle distribution without pitch control (solid curves); the industrial case with optimal active pitch control of the blade as a whole at optimal twist of the blade (dash-dotted curves); the theoretical case with optimal active pitch control of each blade section independently (dashed curves). The first row shows the corresponding evolution of the angles (a) $\phi$, (b) $\theta_0$ and (c) $\beta=\phi-\theta_0$, with respect to the blade radius $r$, for three tip-speed ratios: $\Lambda=3$ (top curves), $\Lambda=\Lambda^{opt}=7.8$ (center curves), $\Lambda=13$ (bottom curves). Pitch control increases $\theta_0$ at low $\Lambda$ to reduce $\beta$ and to avoid stall. Conversely pitch control decreases $\theta_0$ at high $\Lambda$ to adopt a better aerodynamic position. The more degrees of freedom there are, the larger is the wind turbine efficiency.
The second row shows the evolution of (d) the radius-averaged pitch angle $<\theta_0>$, (e) the aerodynamic torque $M$, and (f) the efficiency $C_P$, with respect to the tip-speed ratio $\Lambda$.
Vertical orange dashed lines indicate the tip-speed ratio $\Lambda^{opt}=7.8$ corresponding to the maximal efficiency $C_P^{max}(\Lambda)=0.40$ reached by the turbine.}}
\label{FigU_1}
\end{figure}
\noindent The first step determines the theoretical optimal distribution of the pitch angle along the radius $r$ which maximizes the efficiency $C_P$ for each tip-speed ratio $\Lambda$ of the working range. This function is referred to as $\theta_0^{opt}(r,\Lambda)$. The corresponding optimal efficiency curve is called $C_P^{max}(\Lambda)$.\\
\noindent The angle $\phi=\arctan{\left(R/ (r \Lambda)\right)}$ between the effective wind direction $\mbs{V}$ and the rotation plane is a decreasing function of the radius and of the tip-speed ratio, as shown on figure \ref{FigU_1}(a). At a given value of $\phi$, changing the pitch angle as described by the function $\theta_0^{opt}(r,\Lambda)$ (see Fig. \ref{FigU_1}(b), dashed curves) is a way to optimize the angle of attack of the wind on the blade $\beta_0^{opt}(r,\Lambda)=\phi(r,\Lambda)-\theta_0^{opt}(r,\Lambda)$ (see Fig. \ref{FigU_1}(c), dashed curves). The main effect of optimizing the pitch angle of each blade element independently during operation - which is only theoretical - is to keep the angle of attack $\beta$ in a narrow range, from around $11^{\circ}$ at $\Lambda=3$, far below the optimum working point $\Lambda^{opt}=7.8$, to around $3^{\circ}$ at $\Lambda=13$, far above the optimum working point (see Fig. \ref{FigU_1}(c)). The system avoids stalling at low values of $\Lambda$, increasing significantly the aerodynamic torque and the efficiency. At high values of $\Lambda$, small variations of the pitch angle help the wind turbine to adopt a more aerodynamic position - especially at the blade root - which also increases $M$ and $C_P$. The corresponding variations of the aerodynamic torque $M^{max}(\Lambda)$ and the efficiency $C_P^{max}(\Lambda)$ are shown respectively on figures \ref{FigU_1}(e) and \ref{FigU_1}(f) (dashed curves).\\
\noindent The efficiency curve $C_P^{max}(\Lambda)$ is maximum at $\Lambda^{opt}=7.8$, with $C_P^{max}(\Lambda^{opt})=0.40$ (see Fig. \ref{FigU_1}(f)). This nominal working point corresponds to the pitch distribution called $\theta_0^{opt}(r,\Lambda^{opt})$. By keeping this pitch distribution fixed all along the working range as shown on figure \ref{FigU_1}(b), (solid curve), the variations of $\beta$, $M$ and $C_P$ are found respectively on figures \ref{FigU_1}(c), (e) and (f) (solid curves). This case corresponds to the wind turbine with rigid blades at fixed optimal pitch distribution ($i.e.$ without any pitch control).\\
\noindent Many current wind turbines benefit from active pitch control. Actuators located at the root of the rigid blades are able to achieve simultaneously a solid rotation of all the blades of the turbine around their radial spar. To also mimic this concrete case, we fix the pitch distribution of the blade equal to $\theta_0^{opt}(r,\Lambda^{opt})$ at $\Lambda=\Lambda^{opt}$. Then for each tip-speed ratio $\Lambda$ of the working range, the optimal bodily rotation of the blade which maximizes the efficiency of the turbine is found, but keeping the same twist of the blade. The evolution of the mean value over the radius $r$ of the pitch distribution with respect to $\Lambda$ is displayed on figure \ref{FigU_1}(d) (dash-dotted curve), crossing the line $<\theta_0>=4.4^{\circ}$ at $\Lambda^{opt}=7.8$. The corresponding variations of $\beta$, $M$ and $C_P$ are shown respectively on figures \ref{FigU_1}(c), (e) and (f) (dash-dotted curves). This method is also able to reduce the range of angle of attack $\beta$ experienced by the rigid blade, but it is less efficient than when the pitch angle of every blade element can be varied.\\
\begin{figure}[]
\centering
\includegraphics[height=10cm]{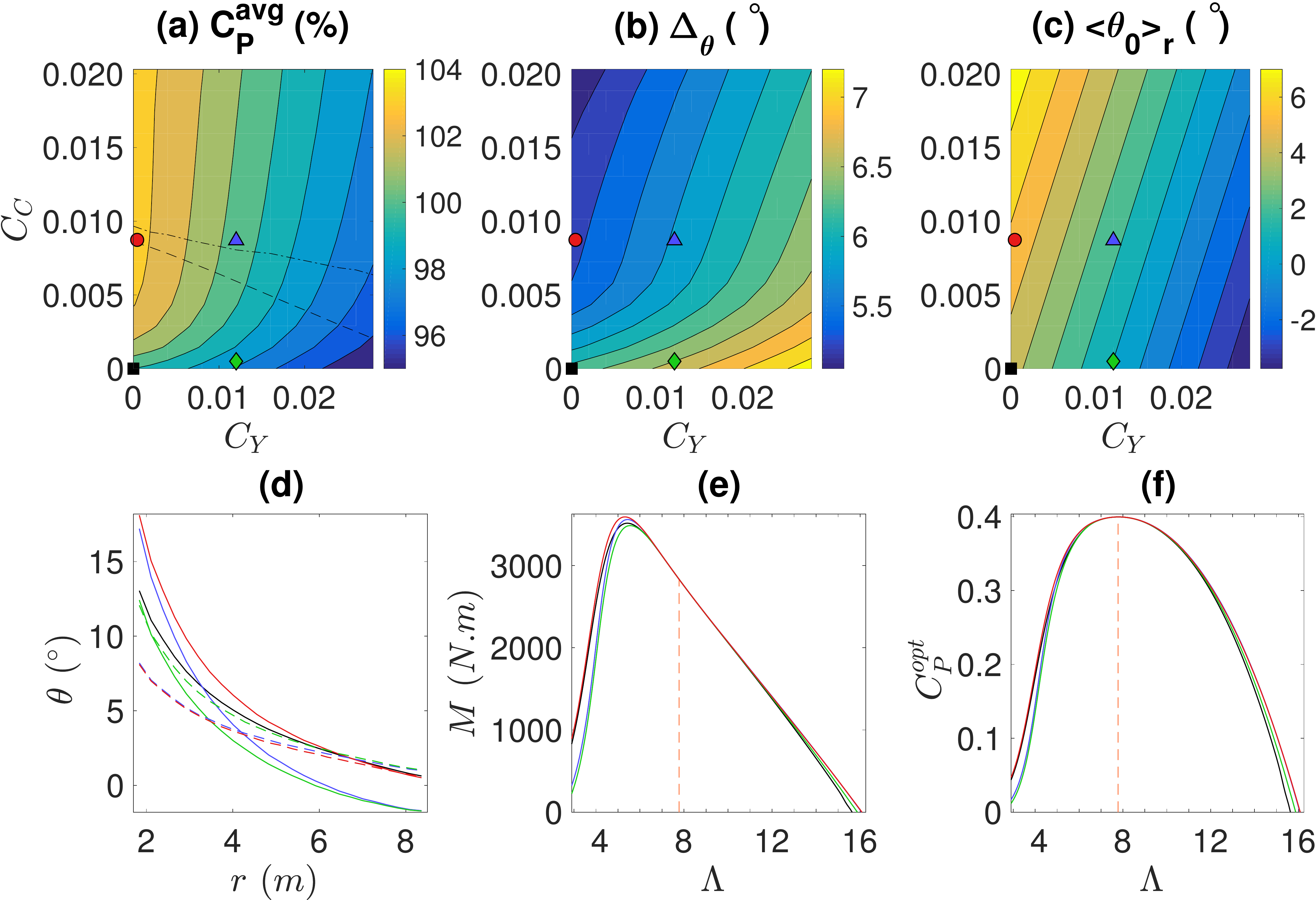} 
\includegraphics[height=10cm]{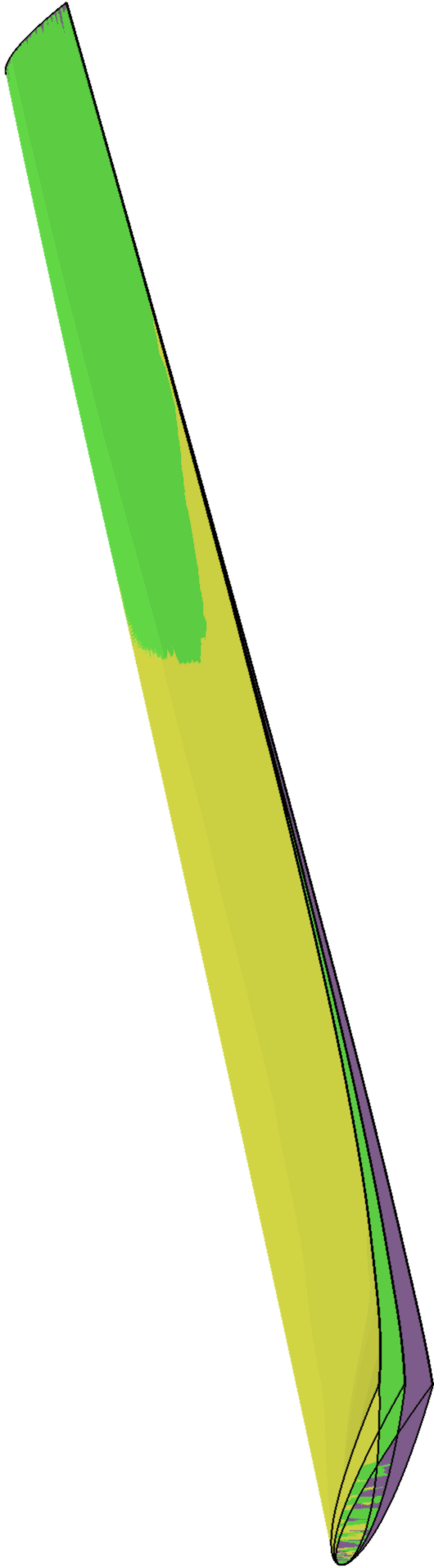} 
\caption{\footnotesize{Evolution of the characteristic parameters of the wind turbine with flexibles blades working at constant incoming wind velocity $U=6.9\ m.s^{-1}$, for a wide range of sets ($C_Y$, $C_C$).
The first row shows the evolution with respect to the Cauchy number $C_Y$ and the centrifugal number $C_C$ of (a) the averaged efficiency $C_P^{avg}$ of the curve $C_P(\Lambda)$, expressed as a percentage of the averaged efficiency corresponding to the rigid case at ($C_Y=0$, $C_C=0$); (b) the radius-averaged deviation $\Delta_{\theta}$ between $\theta(r,\Lambda)$ and the theoretical optimal variation $\theta_0^{opt}(r,\Lambda)$; (c) the radius-averaged initial pitch angle $<\theta_0>_r$ (without any forces) found by the program so that the pitch distribution at $\Lambda^{opt}$ equals $\theta_0^{opt}(r,\Lambda^{opt})$. On figure (a), the dashed curve and the dash-dotted curve correspond respectively to a radius-averaged twist of $1^{\circ}.m^{-1}$ and a maximum twist on the blade of $10^{\circ}.m^{-1}$: the more $C_Y$ and $C_C$ increase, the more the twist increases.
Four sets ($C_Y$, $C_C$) are studied more specifically: (0,0), black square; ($0.5\times 10^{-3}$, $8.7\times 10^{-3}$), red circle; ($12\times 10^{-3}$, $8.7\times 10^{-3}$), blue triangle; ($12\times 10^{-3}$, $0.5\times 10^{-3}$), green diamond.
For these specific cases, the second row shows the corresponding evolution of (d) the pitch distribution $\theta(r)$ at $\Lambda=3$ (solid curves), $\Lambda=\Lambda^{opt}$ (black curve), $\Lambda=13$ (dashed curves); (e) the aerodynamic torque $M$, and (f) the efficiency curves $C_P$ with respect to the tip-speed ratio $\Lambda$.
Vertical orange dashed lines indicate the tip-speed ratio $\Lambda^{opt}=7.8$, corresponding to the maximal efficiency $C_P^{max}(\Lambda)=0.40$ reached by the turbine.
On the right: sketch of three deformations of the flexible blade ($C_Y=0.5\times 10^{-3}$, $C_C=8.7\times 10^{-3}$) at $\Lambda=3.0$ (yellow), $\Lambda=7.8$ (green) and $\Lambda=13.8$ (pink). For comparison, the optimal rigid blade remains in the shape of the green blade over the entire working range.}}
\label{FigU_2}
\end{figure}
\noindent The second step is to find the soft materials for the blades which maximize the averaged efficiency of the wind turbine. Equation \eqref{Eq3} shows that the evolution of the pitch angle $\theta(r,\Lambda)$ is governed by only three parameters: the initial pitch angle distribution $\theta_0(r)$, the Cauchy number $C_Y$ and the centrifugal number $C_C$. To each set ($\theta_0(r)$, $C_Y$, $C_C$) corresponds a unique curve $C_P(\Lambda)$. Since the goal is to optimize the machine around its nominal working point, every curve $C_P(\Lambda)$ is forced to go through the point ($\Lambda^{opt}$, $C_P^{max}(\Lambda^{opt})$), as shown by the vertical orange dashed line on figures \ref{FigU_1}(d-f). This means the function $\theta(r,\Lambda)$ must be equal to $\theta_0^{opt}(r,\Lambda^{opt})$ at tip-speed ratio $\Lambda^{opt}$. This constraint fixes the initial pitch distribution $\theta_0(r)$ for each set ($C_Y$, $C_C$), by following the method displayed on figure \ref{Fig2}.
As displayed in equation \eqref{Eq4}, the Cauchy number and centrifugal number depend on $W^2$ and $W^3$, respectively. Since the chord $W$ varies with the radius, values of both numbers must vary with $r$. To have an idea of their order of magnitude, the values of $C_Y$ and $C_C$ displayed on figures \ref{FigU_2} and \ref{Figf_2} correspond to $W_0=W(R_{min})=1.085\ m$.\\
\noindent Figure \ref{FigU_2}(a) shows the evolution of the mean value of the efficiency curve $C_P(\Lambda)$ with respect to the Cauchy and centrifugal numbers. The mean value $<C_P>=0.23$ of the rigid case, corresponding to $C_Y=0$ and $C_C=0$ (see figure \ref{FigU_1}(f), solid curve), is taken as a reference. In the operating mode at constant $U$, the centrifugal effect increases the mean value of the efficiency curve up to $4\%$, whereas the aerodynamic effect decreases it. Figure \ref{FigU_2}(b) displays the mean value over the tip-speed ratio $\Lambda$ of the deviation $\Delta_{\theta}$ between the pitch angle distribution $\theta_{(C_Y, C_C)}(r, \Lambda)$ and the optimal one $\theta_0^{opt}(r,\Lambda)$ computed at the first step of the method (see Fig. \ref{FigU_1}(b)): 
\begin{equation}
\Delta_{\theta}(C_Y,C_C) = <\sqrt{\frac{\int_{R_{min}}^{R}\left(\theta_{(C_Y, C_C)}(r, \Lambda)-\theta_0^{opt}(r,\Lambda)\right)^2 \mathrm{dr}}{(R-R_{min})}}>_{\Lambda}.
\label{EqDeltaTheta}
\end{equation}
The pattern on this map is linked with that on figure \ref{FigU_2}(a): the overall efficiency of the turbine with flexible blades is all the larger as the pitch variation is close to the optimal one $\theta_0^{opt}(r,\Lambda)$. Figure \ref{FigU_2}(c) shows the evolution of the mean value $<\theta_0(r)>_r(C_Y, C_C)$ of the initial pitch angle distribution. As $C_Y$ increases, the aerodynamic forces tend to increase the mean value of the pitch angle distribution at working point $\Lambda^{opt}$. In order to go through the nominal working point characterized by the pitch distribution $\theta_0^{opt}(r,\Lambda^{opt})$ at $\Lambda^{opt}$, the mean value of the initial pitch angle distribution must decrease. Conversely, as $C_C$ increases, the centrifugal force tends to decrease the mean value of the pitch angle distribution at working point $\Lambda^{opt}$: the mean value of the initial pitch angle distribution must increase. This evolution is independent of the operating mode.\\
\noindent Four sets ($C_Y$, $C_C$) are studied in detail. (0,0) corresponds to the rigid case (black square, $c.f.$ Fig. \ref{FigU_2}(a)). The other cases correspond to different balances between the aerodynamic and centrifugal intensities to see the contribution of each force: (0.5, 8.7) $10^{-3}$ (red circle), (12, 8.7) $10^{-3}$ (blue triangle), and (12, 0.5) $10^{-3}$ (green diamond). The corresponding curves $\theta(r)$ are shown on figure \ref{FigU_2}(d) for different values of the tip-speed ratio: $\Lambda=3$ (solid curves),  $\Lambda=7.8$ (black curve), $\Lambda=13$ (dashed curves). Figures \ref{FigU_2}(e) and (f) display the corresponding evolution of the aerodynamic torque $M(\Lambda)$ and the efficiency $C_P(\Lambda)$, respectively.\\
\noindent The set ($C_Y=0.5\times 10^{-3}$, $C_C=8.7\times 10^{-3}$) (red curve) increases the aerodynamic torque and the efficiency of the turbine for every tip-speed ratio of the working range compared to the rigid case (0,0). At low $\Lambda=3$, the flexible blade is able to passively increase the pitch angle distribution (see Fig. \ref{FigU_2}(d), red solid curve) in order to reduce the stall effect, thus increasing $M$ and $C_P$. At $\Lambda=13$, the flexible blade passively decreases its pitch angle distribution in order to increase the angle of attack $\beta$ and the aerodynamic torque $M$. Because the centrifugal force evolves as $\cos(\theta) \sin(\theta)$ (see Eq. \eqref{Eq3}), this phenomenon is particularly significant at the blade foot where $\theta$ takes the largest values, going from $18.12^{\circ}$ at $\Lambda=3$ to $13.06^{\circ}$ at $\Lambda=7.8$, and to $8.11^{\circ}$ at $\Lambda=13$.\\
\noindent Keeping $C_C=8.7\times 10^{-3}$ while increasing $C_Y$ from $0.5\times 10^{-3}$ to $12\times 10^{-3}$ (blue triangle) only changes the deformation of the blade at small $\Lambda$, where aerodynamic forces dominate the centrifugal force. However this effect is detrimental to the wind turbine efficiency for the following reason. At low tip-speed ratio ($\Lambda \leq \Lambda^{opt}$), the aerodynamic forces tend to increase the pitch angle especially at the tip of the blade where this force is the largest. In order to be equal to $\theta^{opt}(r,\Lambda^{opt})$ at tip-speed ratio $\Lambda^{opt}$, the initial pitch angle at the blade tip must decrease. However, a value of pitch angle which is too small, especially if negative (see Fig \ref{FigU_2}(d), solid blue curve), reduces the aerodynamic torque and performance at start, as shown at low $\Lambda$ on figures \ref{FigU_2}(e,f). At high $\Lambda$, the performance is similar to the previous case corresponding to negligible aerodynamic forces.
In that case where the values of $C_Y$ and $C_C$ are both leading to pitch angle variations of a few degrees in absolute value, we evaluate the quantities $C_Y \left[ (1-a)^2 + (\lambda (1+a'))^2 \right]$ and $C_C \Lambda^2$, which represent the effective aerodynamical and centrifugal intensities compared to the restoring force of the torsion spring, respectively, as shown in Eq. \eqref{Eq3}. At the maximum tip-speed ratio $\lambda=\Lambda=16$, while canceling $a$ and $a'$ to have an order of magnitude, we obtain values up to $3.1$, as announced in part \ref{Modeling}.\\
\noindent Keeping $C_Y=12\times 10^{-3}$ while decreasing $C_C$ from $8.7\times 10^{-3}$ to $0.5\times 10^{-3}$ (green diamond) shows the effect of aerodynamic forces alone. The evolution of the blade tip, where centrifugal force is negligible compared to the aerodynamic forces, is the same as the blue curve over the entire working range. This leads to poor performance at low $\Lambda$ for the same reasons as the previous case. Besides, since the centrifugal effect is turned off, there is almost no deformation at the blade root, which decreases the performance at high and low $\Lambda$ compared to the previous case ($C_Y=12\times 10^{-3}$, $C_C=8.7\times 10^{-3}$) (blue curve).\\
\noindent The largest deformation reached by any blade section over the entire working range for the sets (0.5, 8.7) $10^{-3}$ (red circle), (12, 8.7) $10^{-3}$ (blue triangle), and (12, 0.5) $10^{-3}$ (green diamond) is respectively $13.29^{\circ}$, $11.14^{\circ}$, and $4.13^{\circ}$, allowing us to make the assumption of low deformations. In the same line of thought, the three cases studied have a mean twist smaller or a bit larger than $1^{\circ}.m^{-1}$ (figure \ref{FigU_2}(a), dashed curve) and a maximum twist smaller or a bit larger than $10^{\circ}.m^{-1}$ (figure \ref{FigU_2}(a), dash-dotted curve). As the blade section deformation becomes larger, the curvature of the profile increases and the area exposed to the wind decreases. This effect, which is not taken into account into our model, should reduce the expected performance at high deformations (for instance at very large values of $C_C$).\\
\begin{figure}[]
\centering
\includegraphics[width=1 \linewidth]{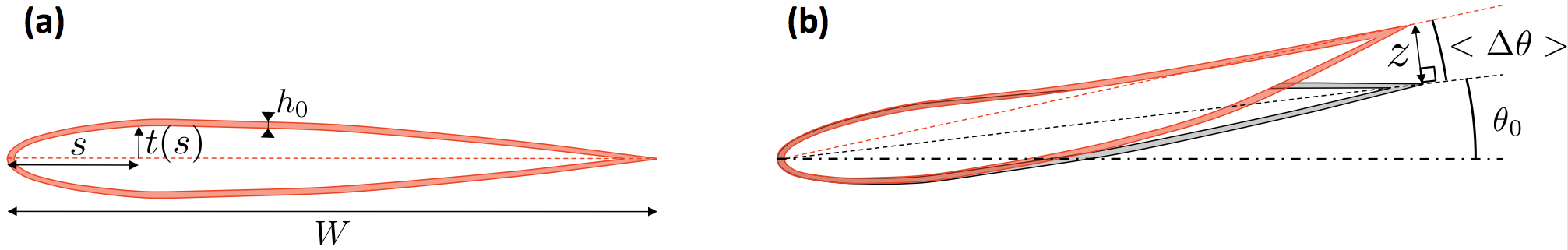} 
\caption{\footnotesize{(a) Empty NACA 0012 airfoil: $t$ is the half-thickness of the profile, $h_0$ is the thickness of the contour (in red) of the Young modulus $E$ and density $\rho_{blade}$. (b) Initial shape (in grey) and deformed shape (in red) of the airfoil, due to aerodynamic and centrifugal forces. $<\Delta \theta>$ is the chord-averaged value of the pitch angle variation.}}
\label{Fig_Step3}
\end{figure}
\begin{figure}[]
\centering
\includegraphics[width=0.6 \linewidth]{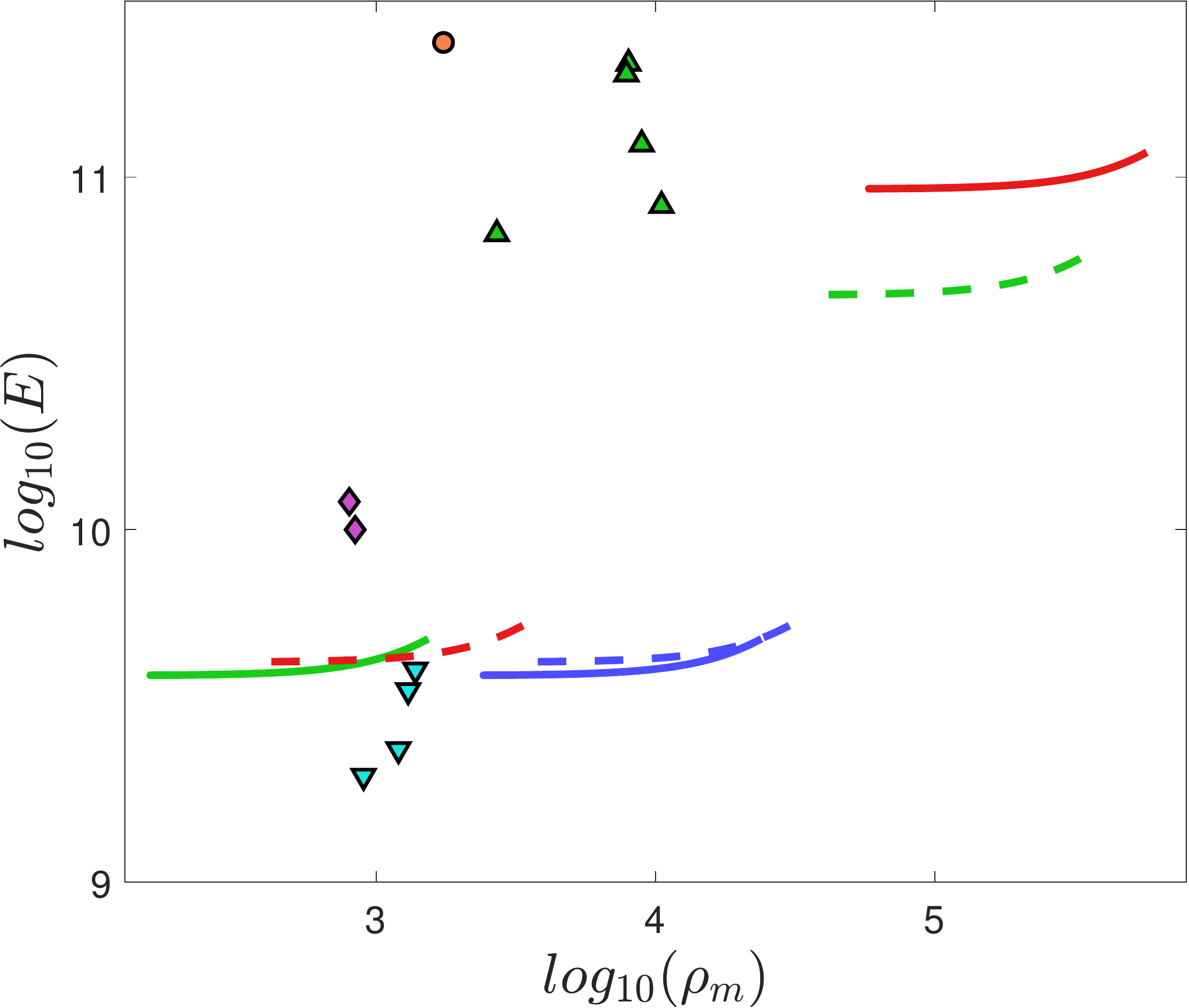} 
\caption{\footnotesize{Log-log evolution of the Young modulus $E$ with respect to the density of the blade contour $\rho_m$ for different dimensionless thicknesses $\tilde{h_0}$ of the contour, from $0.1\%$ to $1\%$ of the chord $W$. Concerning the operating mode at constant $U$, the sets ($C_Y$, $C_C$) displayed are in solid curves: ($0.5\times 10^{-3}$, $7.8\times 10^{-3}$) in red, ($12\times 10^{-3}$,$7.8\times 10^{-3}$) in blue, and ($12\times 10^{-3}$, $0.5\times 10^{-3}$) in green. Concerning the operating mode at constant $\Omega$, the sets displayed are in dashed curves: ($11\times 10^{-3}$, $1\times 10^{-3}$) in red, ($11\times 10^{-3}$, $9\times 10^{-3}$) in blue, and ($1\times 10^{-3}$, $9\times 10^{-3}$) in green. Markers represent existing materials: metals (green uptriangles) like silver, copper, steel, iron and aluminium from right to left; high-strength carbon fiber (orange circle), woods (purple diamonds) like ash tree and oak tree; and polymers (blue downtriangles).}}
\label{Etape3}
\end{figure}
\noindent The third step is to find some practical materials to make the flexible blades. These blades are empty, with a contour of thickness $h_0$, made of a soft and homogeneous material of Young modulus $E$ and density $\rho_{blade}$, as presented on figure \ref{Fig_Step3}(a).
Modeling the blade section deformation with a spring facilitates and speeds up the first two steps. However the use of an homogeneous soft material is more advisable. It has to be noted that the use of non-homogeneous materials could even be more advisable. This issue is not addressed here. Concretely, the goal is to go from an optimal set ($C_Y$, $C_C$) obtained from the model with springs, to some sets ($E$, $\rho_{blade}$, $h_0$) of a empty flexible blade. The beam equation of the blade section was developed in \cite{cognet_bioinspired_2017}, Eq. (3.2). The half-thickness of the profile is referred to as $t$. In order to have a mathematical relation between these two sets, we look for averaged pitch angle variation $<\Delta \theta>(r,\Lambda)$ to be equal in the two models.\\
\noindent In the case of an empty flexible airfoil, at a given radius $r$ and tip-speed ratio $\Lambda$: $<\Delta \theta>=\arctan{\left(z/W\right)}$, where $z$ is the transverse displacement of the trailing edge of the airfoil (see Fig. \ref{Fig_Step3}(b)). Using the assumption of small deformations, $z(W)=\int_{s=0}^{s=W} \mathrm{dz} = \int_0^W \sin(\theta(s)-\theta_0) \mathrm{ds} \approx \int_0^W (\theta(s)-\theta_0) \mathrm{ds}$. Under the same hypothesis, we eventually obtain:
\begin{equation}
<\Delta \theta>=\arctan{\left(\frac{\int_0^W (\theta(s)-\theta_0) \mathrm{ds}}{W}\right)} \approx \int_0^1 (\theta(\tilde{s})-\theta_0) \mathrm{d \tilde{s}},
\label{DeltaTh}
\end{equation}
\noindent where $s=\tilde{s}W$. Integrating the Euler-Bernoulli clamped-free beam equation (3.2) developed in \cite{cognet_bioinspired_2017}, we deduce that:
\begin{eqnarray}
<\Delta \theta>_1  &=  & \int_0^1\mathrm{d \tilde{s}} \int_0^{\tilde{s}}\mathrm{d \tilde{s}'} \int_{\tilde{s}'}^{1}\mathrm{d \tilde{s}''}  \int_{\tilde{s}''}^{1}\mathrm{d \tilde{s}'''} \Bigg(\frac{\rho U^2 W^3}{2 B(\tilde{s}''')} (1+\lambda^2)  \left[C_L(\beta) \cos(\beta) + C_D(\beta) \sin(\beta) \right] \nonumber \\
&   & - \frac{\rho_{blade} U^2 (2h_0) W^4}{R^2 B(\tilde{s}''')} \Lambda^2 \sin(\theta)  \int_0^{\tilde{s}'''}\cos(\theta(\tilde{s}''''))\mathrm{d \tilde{s}''''}\Bigg).
\label{Eq_Beam1}
\end{eqnarray}
\noindent The bending modulus per unit length writes $B(s)=EI(s)$, where $I(s)=2 \int_{t(s)-h_0}^{t(s)} y^2 dy=\frac{2}{3}\left[t^3(s)-(t(s)-h_0)^3\right]$ (if $t(s)-h_0 \leq 0$, then $t(s)-h_0 = 0$ is imposed). Under the assumption of small deformations, we evaluate the right member at $\theta=\theta_0$. Thus $\int_0^{\tilde{s}'''}\cos(\theta(\tilde{s}''''))\mathrm{d \tilde{s}''''}=\tilde{s}''' \cos(\theta_0)$. Most of the terms of the equation no longer depend on $s$, which leads to:
\begin{align}
<\Delta \theta>_1 = J_1 \frac{\rho U^2 W^3}{2 E} (1+\lambda^2)  \left[C_L(\beta) \cos(\beta) + C_D(\beta) \sin(\beta) \right] - J_2 \frac{\rho_{blade} U^2 (2h_0) W^4}{R^2 E} \Lambda^2 \sin(\theta_0) \cos(\theta_0),
\label{Eq_Beam2}
\end{align}
\noindent where $\beta=\arctan{\left(\frac{1}{\lambda}\right)} -\theta_0$, and :
\begin{align}
J_1 = \int_0^1\mathrm{d \tilde{s}} \int_0^{\tilde{s}}\mathrm{d \tilde{s'}} \int_{\tilde{s'}}^{1}\mathrm{d \tilde{s''}}  \int_{\tilde{s''}}^{1} \frac{\mathrm{d \tilde{s'''}}}{I(\tilde{s'''})} \ \ \ ;\ \ \ J_2= \int_0^1\mathrm{d \tilde{s}} \int_0^{\tilde{s}}\mathrm{d \tilde{s'}} \int_{\tilde{s'}}^{1}\mathrm{d \tilde{s''}}  \int_{\tilde{s''}}^{1} \frac{\tilde{s'''} \mathrm{d \tilde{s'''}}}{I(\tilde{s'''})}
\label{}
\end{align}
\noindent Using dimensionless half-thickness $\tilde{t}=\frac{t}{W}$ and dimensionless contour thickness $\tilde{h_0}=\frac{h_0}{W}$, we obtain: 
\begin{align}
\tilde{I}=\frac{I}{W^3}\ \ \ ;\ \ \ \tilde{J_1}=J_1 W^3\ \ \ ;\ \ \ \tilde{J_2}=J_2 W^3.
\label{}
\end{align}
\noindent In the case of the spring-blade-element model presented in the previous part, $<\Delta \theta>(r,\Lambda)=\theta(r,\Lambda)-\theta_0(r)$, which can be computed by using equation \eqref{Eq3}. Taking the loading at $\theta=\theta_0$, and setting $a$ and $a'$ to zero in order to be in the same conditions as equation \eqref{Eq_Beam2}, we obtain:
\begin{align}
<\Delta \theta>_2= C_Y (1+\lambda ^2)  \left(C_L(\beta) \cos(\beta) + C_D(\beta) \sin(\beta) \right) - C_C \Lambda^2 \sin(\theta_0)  \cos(\theta_0).
\label{Eq_Spring1}
\end{align}
\noindent For every radius $r$ and every tip-speed ratio $\Lambda$, the expressions of $<\Delta \theta>$ given in equations \eqref{Eq_Beam2} and \eqref{Eq_Spring1} must be equal. This leads to $C_Y  = (\rho U^2 \tilde{J_1})/(2E) $ and $C_C  = (\rho_{blade} U^2 (2h_0) W  \tilde{J_2})/(E R^2)$. Isolating the Young modulus and the density, these equations become:
\begin{align}
E  = \frac{\rho U^2}{2C_Y} \tilde{J_1} \ \ \ ;\ \ \ \rho_{blade}  = \rho  \frac{C_C}{C_Y} \frac{R^2}{2h_0 W} \frac{\tilde{J_1}}{\tilde{J_2}},
\label{Eq_ERho1}
\end{align}
\noindent where $W=W_0$ in equation \eqref{Eq_ERho1} as explained at the second step. The solutions of the system \eqref{Eq_ERho1} are displayed on figure \ref{Etape3} in solid curves, with the corresponding colors shown in figure \ref{FigU_2}, for thickness of the contour $\tilde{h_0}$ between $0.1\%$ and $1\%$ of the chord 
The right end of each curve correspond to smallest thickness $h_0$ of the contour, where there are high density $\rho_{blade}$ and high Young modulus $E$.\\
\noindent The soft material solution to build the flexible blade shown on figure \ref{Etape3} are close to materials that already exist (in markers). The soft materials corresponding to the set ($C_Y=0.5\times 10^{-3}$, $C_C=8.7\times 10^{-3}$), in red solid curve, are denser than the usual metals shown in green triangles directed upward. One way to use metals for this case is to increase the thickness $h_0$ of the contour. However this method also increases the total amount of material used, so a compromise must be found. Concerning the other cases ($C_Y=12\times 10^{-3}$, $C_C=8.7\times 10^{-3}$), in blue solid curve, and ($C_Y=12\times 10^{-3}$, $C_C=0.5\times 10^{-3}$), in green solid curve, polymers (blue triangles directed downward) can both be used with values of $\tilde{h_0}$ equal to $1.74\%$ and $0.24\%$, respectively.
\section{Discussion}
\begin{figure}[]
\centering
\includegraphics[width=1 \linewidth]{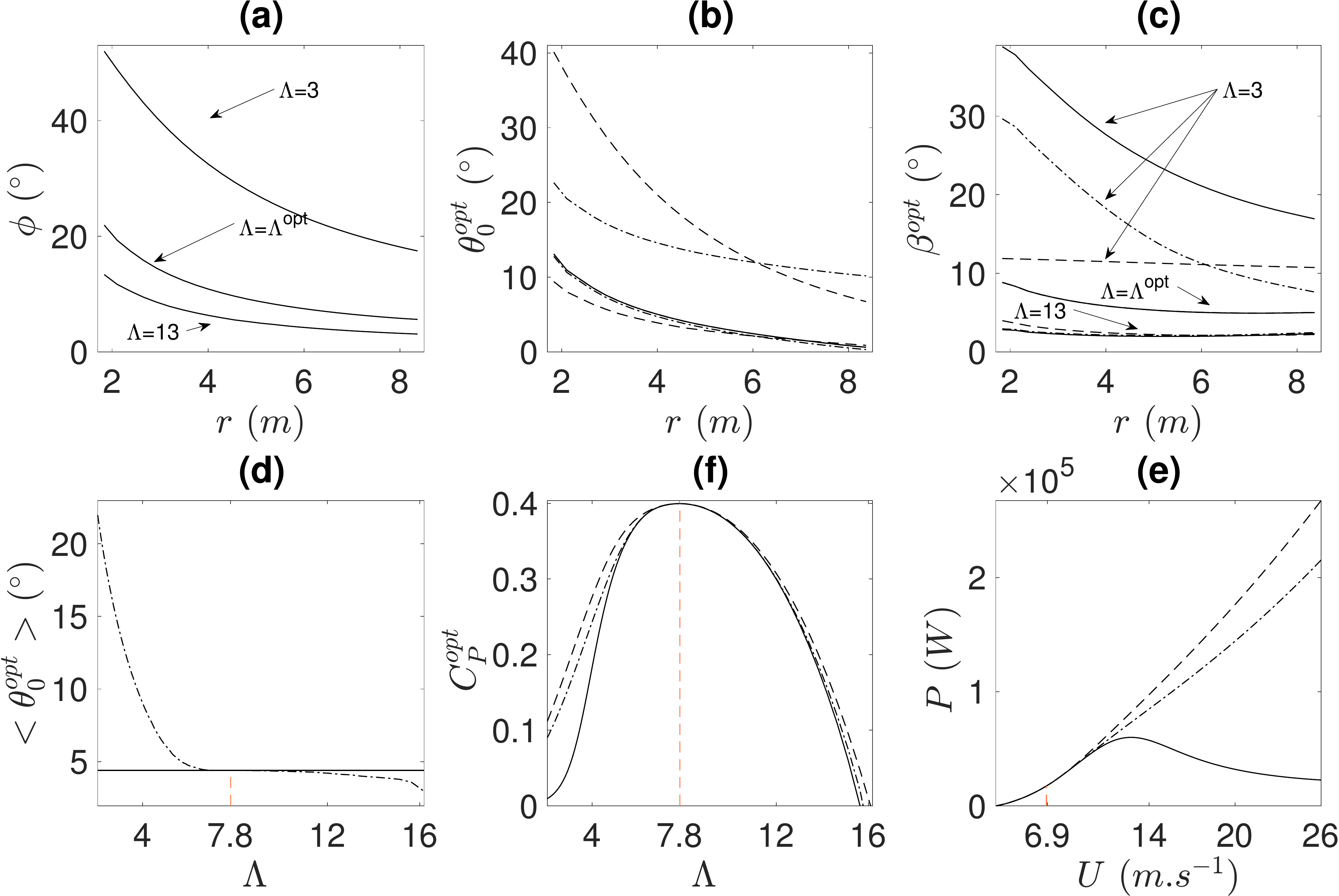} 
\caption{
\footnotesize{Evolution of the characteristic parameters of the wind turbine working at constant rotation rate $f=1\ Hz$, for three types of pitch angle variations: the reference case at optimal fixed pitch angle distribution, without pitch control (solid curves); the industrial case with optimal active pitch control of the blade as a whole, at optimal twist of the blade (dash-dotted curves); the theoretical case with optimal active pitch control of each blade section independently (dashed curves). The first row shows the corresponding evolution of the angles (a) $\phi$, (b) $\theta_0$ and (c) $\beta=\phi-\theta_0$, with respect to the blade radius $r$, for three tip-speed ratios: $\Lambda=3$ (top curves), $\Lambda=\Lambda^{opt}=7.8$ (center curves), $\Lambda=13$ (bottom curves). Pitch control increases $\theta_0$ at low $\Lambda$ to reduce $\beta$ and to avoid stall. Conversely pitch control decreases $\theta_0$ at high $\Lambda$ to adopt a more aerodynamic position. The more degrees of freedom there are, the larger the wind turbine efficiency is.
The second row shows the evolution of (d) the radius-averaged pitch angle $<\theta_0>$, (e) the efficiency $C_P$, and (f) the rotational power $P$, with respect to the tip-speed ratio $\Lambda$. Pitch control is crucial at high values of $\Lambda$.
Vertical orange dashed lines indicate the tip-speed ratio $\Lambda^{opt}=7.8$ corresponding to the maximal efficiency $C_P^{max}(\Lambda)=0.40$ reached by the turbine.}}
\label{Figf_1}
\end{figure}
In this section, we discuss the effects of size, operating mode, and wind velocity distribution on the results of this method.\\

To test the size effect, this method has been applied to wind turbines with the same geometry and same blade profile as the one above, but with all lengths ($W$, $R$, $t$, $h_0$) multiplied by a factor $\alpha$. If the wind velocity $U$ is still equal to $6.9\ m.s^{-1}$, the results show that the efficiency curve $C_P(\Lambda)$, the angle curves $\phi(r,\Lambda)$, $\beta(r,\Lambda)$, $\theta(r,\Lambda)$, and the Cauchy and centrifugal numbers are all unchanged. The new torque $M'$, the new power $P'$ and the new twist angle per unit length $t_w'$ resulting from this scaling are changed as follow :
\begin{align}
M'=M \alpha^3,\ \ \ P'=P \alpha^2,\ \ \ t_w'=t_w\ \alpha^{-1}.
\label{Alpha}
\end{align}
The results on the torque and power from equation \eqref{Alpha} are well known \cite{burton_wind_2011, wood_blade_2001}. Similarly, the result on $t_w'$ is straightforward since the pitch angle distribution is unchanged whereas the radius is multiplied by $\alpha$. However, the fact that the optimal set ($C_Y$, $C_C$) does not vary with size is new and has significant consequences; from equation \eqref{Eq_ERho1}, we conclude that the optimal set of Young modulus $E$ and density $\rho_m$ does not vary with the size of the turbine either. The optimal soft materials found on a model in a wind tunnel are theoretically scale-independant. Experiments on small wind turbines are a possible way to determine the optimal soft materials for larger wind turbines.\\
The model does not take into account the effect of gravity which becomes increasingly significant as size increases. In addition, the wind velocity received by the turbine usually increases with size, for example because of the atmospheric boundary layer. For a different incoming wind velocity $U$, the method will give another optimal soft material. In the same line of thought, changing the profile of the blade or the Reynolds number modifies the aerodynamic coefficients \cite{guyon_hydrodynamique_2012} and the set ($E$, $\rho_m$) of the optimal soft materials.\\

To test the impact of the operating mode, this method has been conducted at constant rotation rate $f$. The geometry and the size of the wind turbine are the same as those presented in part \ref{Method.}. In order to keep the same optimal point obtained at constant incoming wind velocity $U=6.9\ m.s^{-1}$ corresponding to ($\Lambda^{opt}=7.8$, $f^{opt}=1\ Hz$, $C_P^{max}=0.40$), we fixed the rotation frequency $f=1\ Hz$, so that when $U$ equals $6.9\ m.s^{-1}$, the efficiency $C_P$ is maximum and equal to 0.40.\\
For this operating mode, equation \eqref{Eq3} shows that $C_C \Lambda^2=\rho_{blade}(2\pi f)^2 h_0 W^3 dr/(3K)$ does not vary when $f$ is kept constant. The only way for the centrifugal force to vary is for aerodynamic forces to change the pitch angle $\theta$, which will modify the part $\sin(\theta) \cos(\theta)$ in the centrifugal term. However this effect remains small compared to the direct effect of aerodynamic force. At constant rotation rate $f$, centrifugal force distribution still varies with the radius $r$, but almost does not change all along the working range.\\
\begin{figure}[]
\centering
\includegraphics[height=9.7cm]{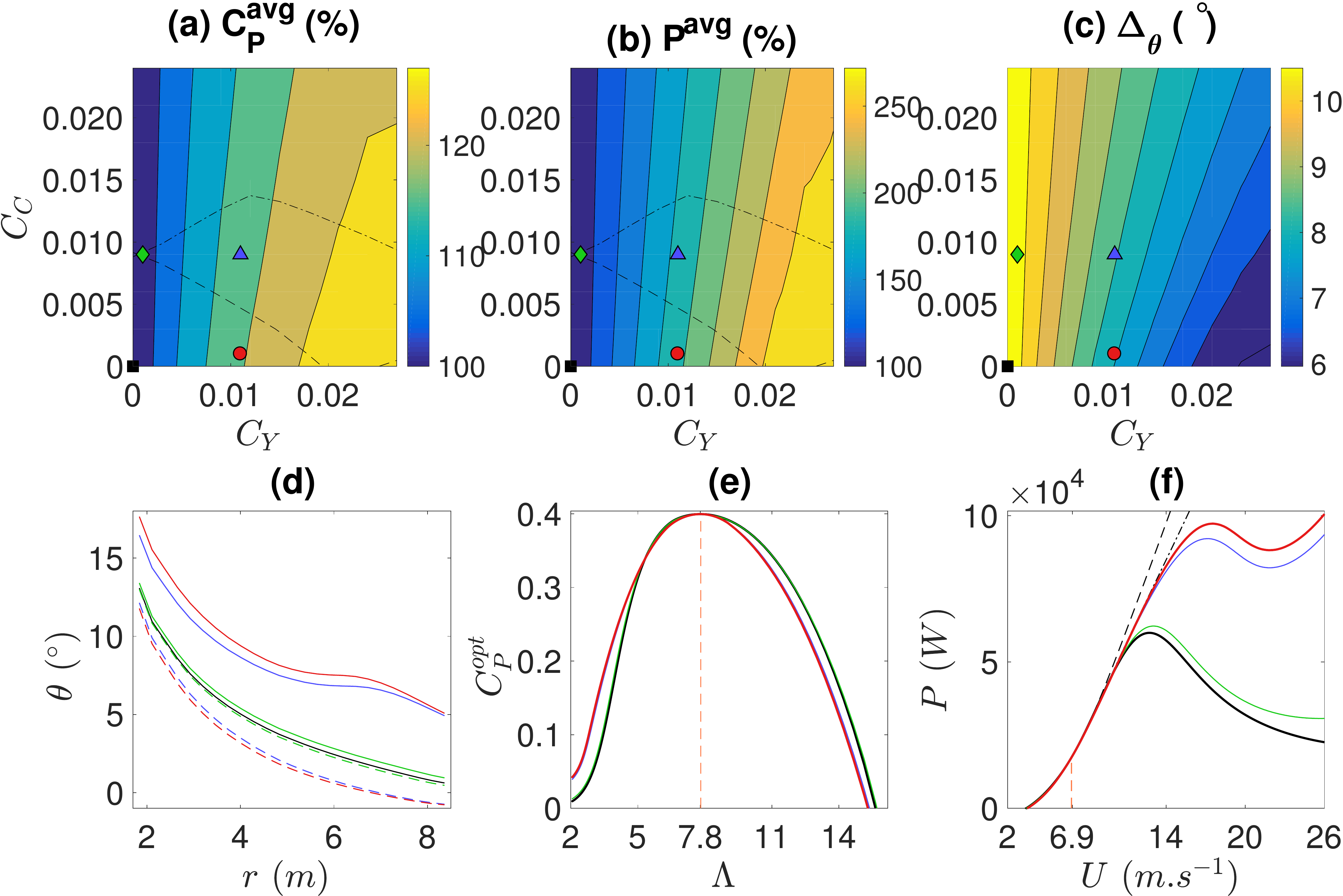} 
\includegraphics[height=9.7cm]{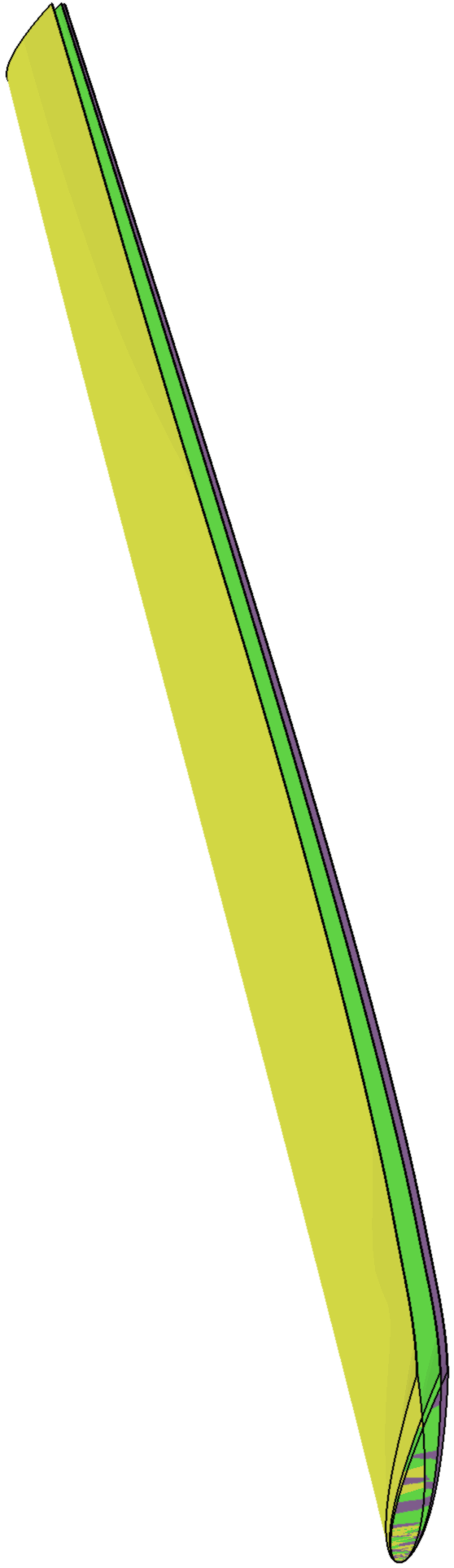} 
\caption{\footnotesize{Evolution of the characteristic parameters of the wind turbine with flexible blades working at constant rotation rate $f=1\ Hz$, for a wide range of sets ($C_Y$, $C_C$).
The first row shows the evolution with respect to the Cauchy number $C_Y$ and the centrifugal number $C_C$ of (a) the averaged efficiency $C_P^{avg}$ of the curve $C_P(U)$, and (b) the averaged power of the curve $P(U)$, both expressed as a percentage of the averaged efficiency/power corresponding to the rigid case at ($C_Y=0$, $C_C=0$); (c) the radius-averaged deviation $\Delta_{\theta}$ between $\theta(r,\Lambda)$ and the theoretical optimal variation $\theta_0^{opt}(r,\Lambda)$. On (a,b), the dashed curve and the dash-dotted curve correspond to a radius-averaged twist of $1^{\circ}.m^{-1}$ and a maximum twist on the blade of $10^{\circ}.m^{-1}$ respectively. The more $C_Y$ and $C_C$ increase, the more the twist increases.
Four sets ($C_Y$, $C_C$) are studied more specifically: (0,0), black square; ($11\times 10^{-3}$, $1\times 10^{-3}$), red circle; ($11\times 10^{-3}$, $9\times 10^{-3}$), blue triangle; ($1\times 10^{-3}$, $9\times 10^{-3}$), green diamond.
For these four cases, the second row shows the corresponding evolution of (d) the pitch distribution $\theta(r)$ at $\Lambda=3$ (solid curves), $\Lambda=\Lambda^{opt}$ (black curve), $\Lambda=13$ (dashed curves); (e) the efficiency curves $C_P$, and (f) the power $P$, with respect to the tip-speed ratio $\Lambda$.
Vertical orange dashed lines indicate the tip-speed ratio $\Lambda^{opt}=7.8$, corresponding to the maximal efficiency $C_P^{max}(\Lambda)=0.40$ reached by the turbine.
On the right: sketch of three deformations of the flexible blade ($C_Y=11\times 10^{-3}$, $C_C=1\times 10^{-3}$) at $\Lambda=3.0$ (yellow), $\Lambda=7.8$ (green) and $\Lambda=13.8$ (pink). For comparison, the optimal rigid blade remains in the shape of the green blade over the entire working range.
}}
\label{Figf_2}
\end{figure}
\noindent As in part \ref{Method.}, the three steps of the method are presented. Figure \ref{Figf_1}(a,b,c) illustrated the variations of the angles $\phi(r)$, $\theta_0^{opt}(r)$ and $\beta^{opt}(r)$ at different tip-speed ratio $\Lambda$ equal to 3, 7.8 and 13. The same three types of pitch control for rigid blades as previously are examined: the theoretical optimal blade which can change its radial pitch angle distribution at each working point $\Lambda$ (dashed curves); the blade at fixed optimal pitch angle, without pitch control device (solid curve); and the blade at fixed optimal pitch twist, with the optimal solid rotation of the three blades together (dash-dotted curves). The evolution of the mean pitch angle over the radius with respect to $\Lambda$ is shown on figure \ref{Figf_1}(d) in dash-dotted curves.\\
Similarly to the results obtained for a constant incoming wind velocity, the efficiency $C_P(\Lambda)$ (see Fig. \ref{Figf_1}(e)) is all the larger as the degrees of freedom are numerous. Figure \ref{Figf_1}(f) shows the corresponding power curve $P(U)$. Because $f$ is now constant and $U$ varies, the efficiency curve and power curve are very different for this operating mode. First the tip-speed ratio decreases with the incoming wind velocity, so that high values of $\Lambda$ correspond to low values of $U$, and low values of $\Lambda$ correspond to high values of $U$. Second, because of the term $U^3$ in equation \eqref{Eq6}, small efficiency gain at low $\Lambda$ means large increase of power at high $U$, whereas efficiency gain at high $\Lambda$ gives small power gain at low $U$.\\
\noindent Figure \ref{Figf_2} displays the results of the second step. Contrary to the results obtained for the previous operating mode, optimal sets ($C_Y$, $C_C$) for efficiency gain and power gain (see Fig. \ref{Figf_2}(a,b)) are now located in regions of small ratios between centrifugal and Cauchy numbers. The efficiency gain reaches $+30\%$ whereas the power gain goes up to $+150\%$. This difference between efficiency and power gains is due to the multiplication factor $U^3$ between $C_P$ and $P$. As for constant incoming wind velocity, the deviation map displayed on figure \ref{Figf_2}(c) between the pitch distribution obtained with flexible blades and the optimal pitch distribution computed at step one is strongly related to the efficiency and power gain: low values of $\Delta_{\theta}$ increases the performance of the turbine.\\
In addition to the rigid case (0,0), three sets ($C_Y$, $C_C$) are investigated corresponding to flexible cases. The set ($11\times 10^{-3}$, $1\times 10^{-3}$) is shown on Fig.\ref{Figf_2}(a) by a red circle. The remaining centrifugal force is turned off so that the pitch angle variation is only due to aerodynamic force (see Fig.\ref{Figf_2}(d), red curves). The pitch angle decreases from $\Lambda=3$ (red solid curve) to $\Lambda=13$ (red dashed curve) all along the radius. The efficiency curve $C_P(\Lambda)$ is better than its rigid counterpart at small tip-speed ratio, but has lower performance at high tip-speed ratio (see Fig.\ref{Figf_2}(d), red curve), so that on average the total efficiency does not change substantially. Because of the multiplication factor $U^3$ mentioned earlier, the total power gain exposed on figure \ref{Figf_2}(f) as the red curve for that set is $+94.5\%$.\\
Keeping the aerodynamic force on ($C_Y=11\times 10^{-3}$), the fact that we increase the centrifugal number $C_C$ up to $9\times 10^{-3}$ (blue triangle on Fig.\ref{Figf_2}(a,b), blue curves on Fig.\ref{Figf_2}(d,e,f)) almost does not change the pitch angle variation nor the performance of the wind turbine. The centrifugal force is already nearly constant because of the choice of this operating mode. The total power is increased by $+85.7\%$.\\
Eventually, decreasing the aerodynamic force to $C_Y=1\times 10^{-3}$ while maintaining $C_C=9\times 10^{-3}$ means almost no deformation of the blade (green diamond on Fig.\ref{Figf_2}(a,b), green curves on Fig.\ref{Figf_2}(d,e,f)). As a result, the pitch angle variation and the performance of this set are close to those of the rigid case (black square, black curves). The total power is increased by $+8.8\%$.
In the operating mode at constant rotation rate, the centrifugal force variations with $\Lambda$ are negligible. The results above show it is not possible to improve the performance of the rigid case over the entire working range by using flexible blades. However it is possible to increase significantly the total efficiency of the machine, and the total power to a larger degree. The three sets ($11\times 10^{-3}$, $1\times 10^{-3}$), ($11\times 10^{-3}$, $9\times 10^{-3}$) and ($1\times 10^{-3}$, $9\times 10^{-3}$) studied previously reached a maximal pitch angle variation on the working range equal to $12.5^{\circ}$, $10.0^{\circ}$ and $8.0^{\circ}$ respectively. The mean and the maximal twist of these three flexible cases are always smaller or slightly larger than $1^{\circ}.m^{-1}$ and $10^{\circ}.m^{-1}$ respectively.\\
\noindent Using equations \eqref{Eq_ERho1}, the corresponding sets of solutions ($E$, $\rho_m$) obtained are displayed on figure \ref{Etape3} in dashed curves in the respective color of their markers, for dimensionless thickness $\tilde{h_0}$ of the contour of the profile varying from $0.1\%$ to $1\%$. The right end of each curve corresponds to the smallest thickness $\tilde{h_0}$ of the contour, where there are high density $\rho_{blade}$ and high Young modulus $E$.
To improve significantly the performance of the turbine, it is possible to find a material corresponding to the set ($C_Y=11\times 10^{-3}$, $C_C=1\times 10^{-3}$) by using polymers of Young modulus $E=4.4\times 10^9\ N.m$, density $\rho_m=1.38\times 10^3\ kg.m^{-3}$ and a dimensionless contour thickness $\tilde{h_0}=0.24\%$. Laminates are able to reach such properties. More simply, mylar (Polyethylene terephthalate) can be used for small wind turbines. The weight of this blade of radius $R=8.5\ m$ is $35\ kg$. Assuming that the mass of the blade increase as the cube of the radius, for $R=45\ m$ and $R=80\ m$, the weight of the blade would be 5.2 tons and 29 tons, respectively (ie between $5\%$ and $20\%$ lighter than the current rigid blades).\\ 
\begin{figure}[]
\centering
\includegraphics[width=1 \linewidth]{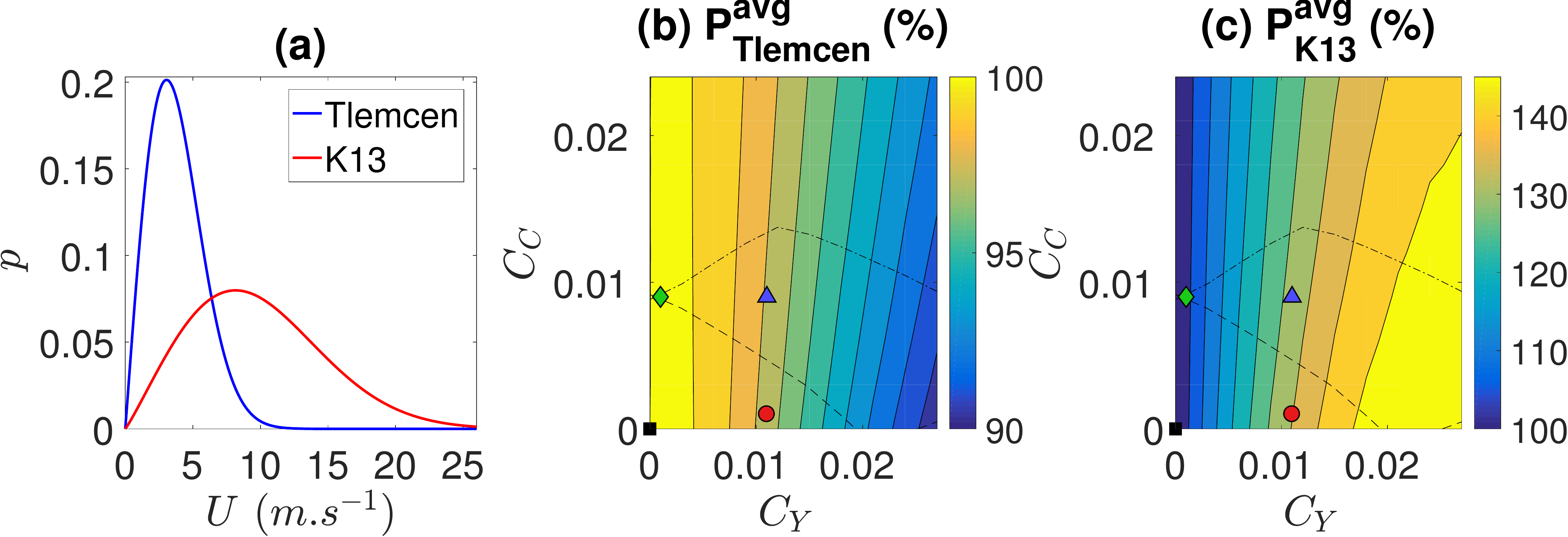} 
\caption{\footnotesize{(a) Probability density curves of incoming wind velocity $U$ in Tlemcen (Algeria), and on platform K13 in the North Sea. (b,c) Maps of averaged power obtained in the case of constant rotation rate $f=1\ Hz$, with respect to the Cauchy and centrifugal numbers, weighted by the probability densities in Tlemcen and on platform K13 respectively, and expressed as a percentage of their corresponding rigid case.
}}
\label{DistributionsEtCartesPuissanceTlemcenK13}
\end{figure}

To test the importance of the wind velocity distribution, we have taken into account two types of distributions as displayed on figure \ref{DistributionsEtCartesPuissanceTlemcenK13}(a).
The first one was measured in Tlemcen, Algeria (solid blue curve). It is fitted by a Weibull characteristic function of shape parameter $k=2.02$ and scale parameter $A=4.29$ \cite{maouedj_etude_2007}. This distribution is narrow and centered around low incoming wind velocities. The second distribution was measured on the platform K13, in the North Sea (red solid curve). It is also fitted by a Weibull function of shape parameter $k=2.1$ and scale parameter $A=11.1$ \cite{Coelingh_Analysis_1996}.
This distribution is wide and takes non negligible values up to $U=26\ m.s^{-1}$. We took this distribution to test the improvement of versatility of turbines with flexible blades. To deduce the wind velocity distribution at the altitude of $2R=17\ m$ from the distribution at the altitude of $74.8\ m$ above the mean sea level given in \cite{Coelingh_Analysis_1996}, we used the logarithmic wind profile: $U(z) \propto \log(z/z_0)$, where $z$ is the altitude above mean sea level, and $z_0$ is the roughness length, equal to 0.0002 in the open sea.\\
Figure \ref{DistributionsEtCartesPuissanceTlemcenK13}(b) and (c) show the power maps with respect to the Cauchy and centrifugal numbers of the wind turbine studied previously (see Fig.\ref{Figf_2}(b)), weighted respectively by the Tlemcen and K13 incoming wind velocity distributions. Since the distribution from Tlemcen amplifies the performance at low $U$ and neglects that at high $U$, the optimal set remains the rigid case (0,0). In contrast, the distribution from K13, which is closer to distributions in real sites of implementation, keeps the same pattern as the one on figure \ref{Figf_2}(b). Total power increase for the sets ($C_Y=11\times 10^{-3}$, $C_C=1\times 10^{-3}$), ($C_Y=11\times 10^{-3}$, $C_C=9\times 10^{-3}$), ($C_Y=1\times 10^{-3}$, $C_C=9\times 10^{-3}$) are $+35\%$, $+32\%$ and $+4\%$, respectively. We repeated the same study for different stations in the North Sea like the Euro Platform (EPF) and Measuring Post Noordwijk (MPN) \cite{Coelingh_Analysis_1996}. The total power gain at ($C_Y=11\times 10^{-3}$, $C_C=1\times 10^{-3}$) are $+24\%$ and $+21\%$, respectively.
\section{Conclusion}
We have presented a general method based on a universal scaling to determine the optimal soft materials for the flexible blades of wind turbines with required geometry and operating mode. Using flexible blades, which are between $5\%$ and $20\%$ lighter than their rigid counterparts, is a way to passively increase the total efficiency of horizontal-axis wind turbines operating at constant wind velocity or at constant rotation rate. We found a significant increase of the total harvested power, varying from $+4\%$ to $+150\%$, depending on the operating mode. Taking into account realistic incoming wind velocity distributions, the power increase still goes up to $+35\%$.\\
Other preliminary results lead us to believe that this method can be adapted for any other operation modes : constant resistive torque, constant tip-speed ratio, resistive torque proportional to the rotation speed, etc.
In this article, we maximized the total efficiency of the turbine, by forcing the efficiency curve to go through the working point corresponding to the maximal efficiency. This scaling can also be applied to reach other goals, such as for instance, to reduce the minimal incoming wind velocity required to start the wind turbine.\\
Finally, because the aerodynamic torque, the elastic restoring torque and the centrifugal torque on the blade all grow as the cube of the characteristic size of the blade, the optimal soft material found for a given wind turbine geometry is, within the limits of the assumptions of our model, scale-independent. Thus, experimental results on small models in wind tunnels are accurate to determine the optimal soft material for larger wind turbines under the same experimental conditions.
\section{Appendix: the BEM theory for flexible blades}
\label{AppendixBEM}
\begin{figure}[]
\centering
\includegraphics[width=1 \linewidth]{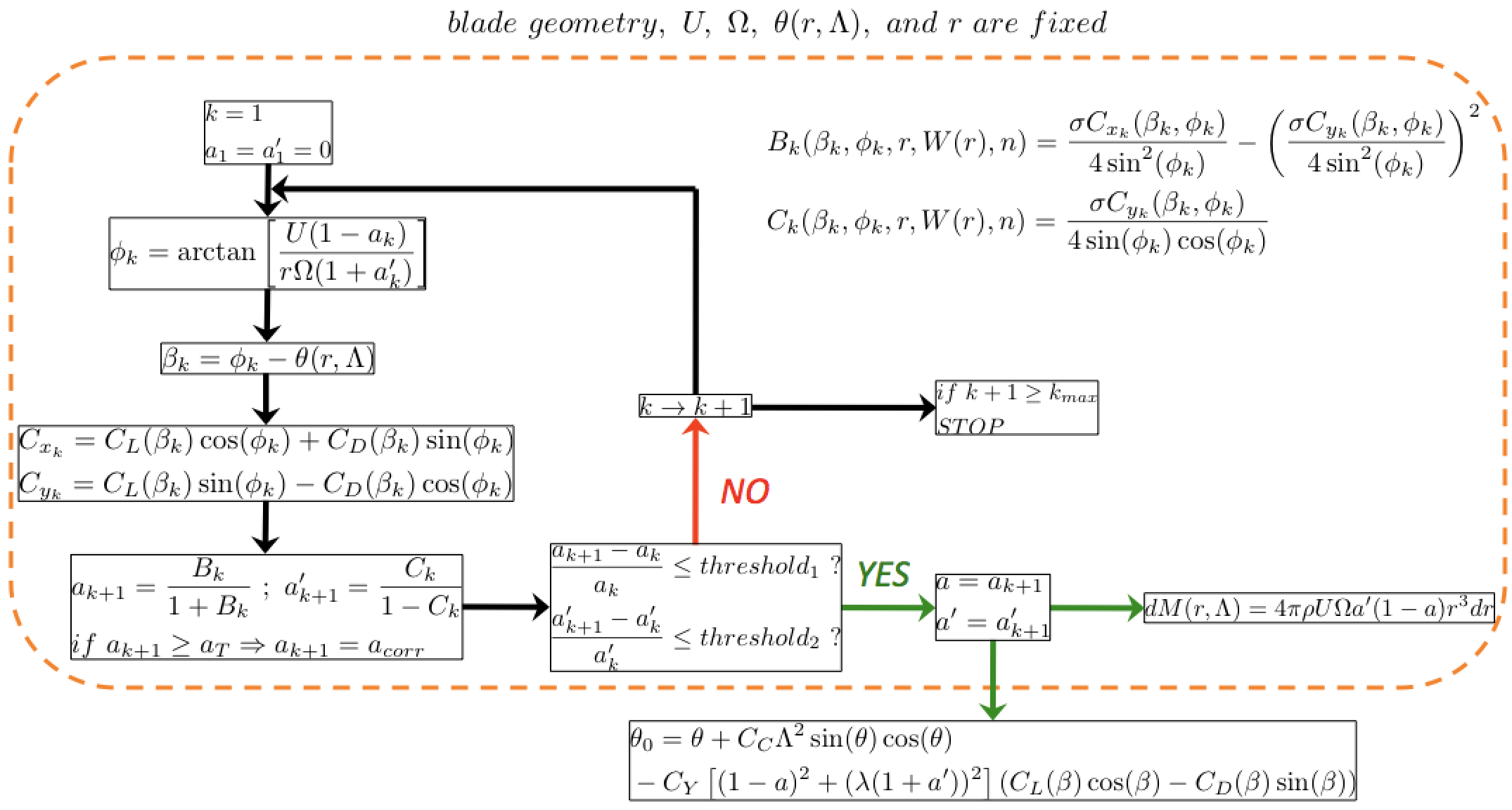} 
\caption{\footnotesize{In the orange dashed rectangle: diagram explaining the steps for finding the axial induction coefficient, $a$, and the tangential induction coefficient, $a'$, from the BEM theory on a blade section located at radius $r$. The incoming wind velocity $U$, the angular velocity $\Omega$, the pitch angle $\theta(r,\Lambda)$ are fixed. At step one, $a$ and $a'$ are initialized to zero (others initial values are possible to speed up the procedure, depending on the operating point), and the BEM theory is applied. At each step $k+1$ of the procedure, the resulting value is compared to the initial guess obtained at step $k$. If the error is less than a threshold (less than a $0.2\%$ for example), the procedure is said to converge. The aerodynamic torque $dM(r,\Lambda)$ of the blade section is computed. Below the orange dashed rectangle: the initial pitch angle $\theta_0$ of the flexible blade section can be easily found from the previous calculation by using equation \eqref{Eq3}.}}
\label{Fig2}
\end{figure}
\noindent The procedure in the orange dashed rectangle on figure \ref{Fig2} presents a common way to solve the equations of the BEM theory \cite{lanzafame_fluid_2007,Sanderse_Aerodynamics_2009}. See \cite{shen_tip_2005, crawford_re-examining_2006} for historical evolution of the BEM theory and its corrections. For a given blade section located at a radius $r$ from the center, with a given geometry, incoming wind velocity $U$, angular velocity $\Omega$, and a given pitch angle $\theta$, the induction coefficients and the local aerodynamic torque $dM(r,\Lambda)$ are computed (orange dashed rectangle). By applying this computation on a blade element at each radius $r$, the total aerodynamic torque $M(\Lambda)=\int_{r}dM(r,\Lambda)$ on the rotor is deduced for any tip-speed ratio of the working range. Since the study concerns the steady-state, the total aerodynamic torque $M$ is always equal to the resistive torque $C$.
Once the induction coefficients are found, the spring equation \eqref{Eq3} can be solved. On figure \ref{Fig2}, the initial pitch angle $\theta_0(r)$ of the flexible blade section is computed from the value of the pitch angle $\theta(r,\Lambda)$ at a given working point (see equation below the orange dashed rectangle).
\newpage
\section{Acknowledgements}
We would like to thank X. Benoit Gonin for helping us to make figures on CATIA, and L. Tuckerman and J.H.A. Cognet for their careful reading of the manuscript. This project was funded by ANR-10-IDEX-0001-02 PSL, from program "Investissements d'avenir".
%
\bibliography{Article_V6s}

\end{document}